\documentclass[twocolumn,showpacs,prb,preprintnumbers,amsmath,amssymb]{revtex4}

\usepackage{graphicx,graphics}
\usepackage{dcolumn}
\usepackage{bm}
\usepackage{epsfig}

\begin{document}

\title{
Fine structure of the local pseudogap and Fano effect\\ 
for superconducting electrons near a zigzag graphene edge 
}

\author{Grigory Tkachov}
\affiliation{
Max Planck Institute for the Physics of Complex Systems, 01187 Dresden, Germany
}


\begin{abstract}
Motivated by recent scanning tunneling experiments 
on zigzag-terminated graphene 
this paper investigates an interplay of evanescent 
and extended quasiparticle states in 
the local density of states (LDOS) near a zigzag edge 
using the Green's function of the Dirac equation. 
A model system is considered where the local electronic structure 
near the edge influences transport of both normal 
and superconducting electrons via a Fano resonance. 
In particular, the temperature enhancement of the 
critical Josephson current and $0-\pi$ transitions are predicted. 
\end{abstract}


\maketitle

\begin{figure}[b]
\epsfxsize=0.8\hsize
\epsffile{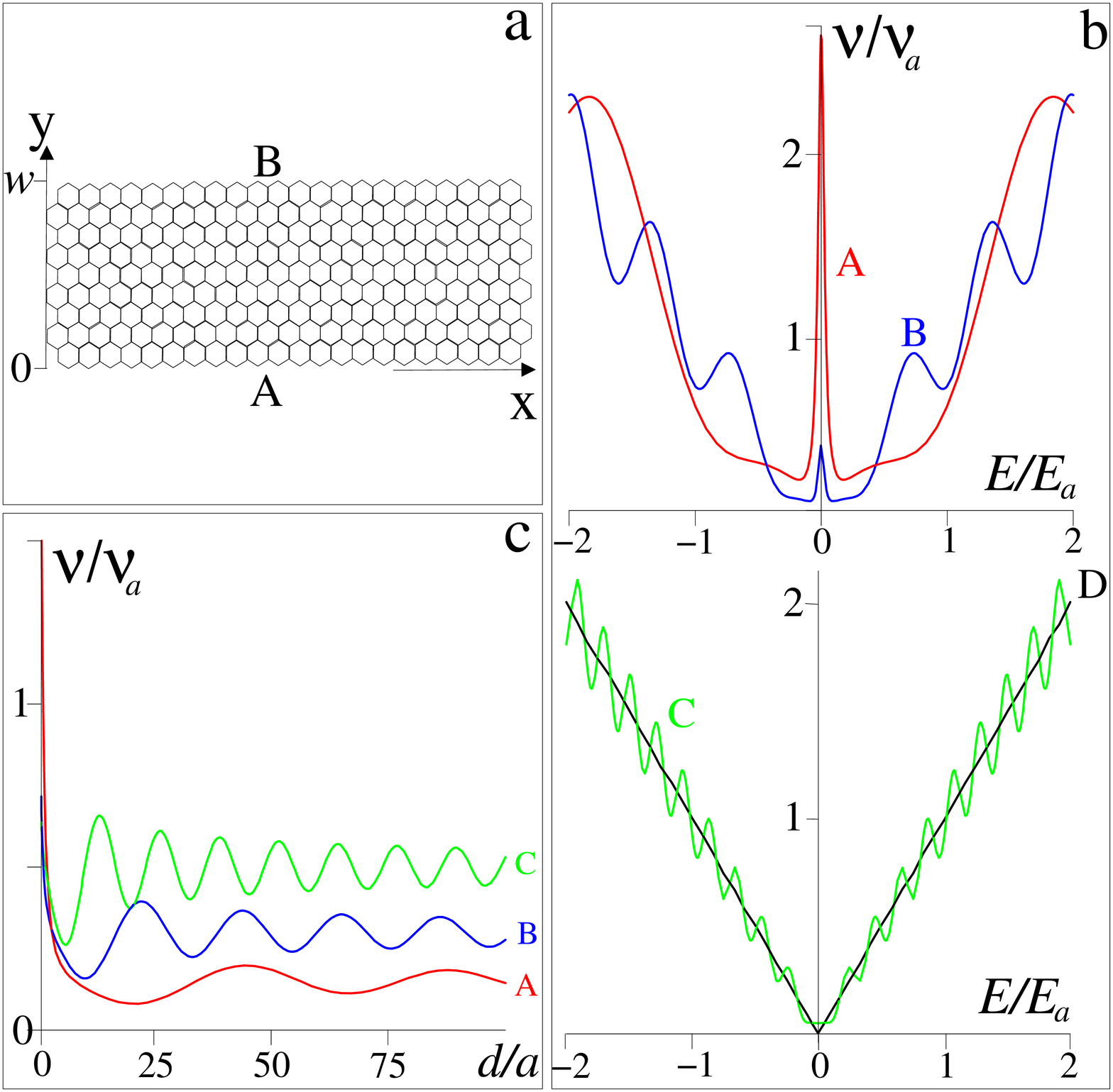}
\caption{(Color online)
(a) Schematic view of a zigzag graphene ribbon 
terminated by atomic lines belonging to different sublattices, 
conventionally denoted as A and B.   
(b)
LDOS vs. energy at different distances from the edge: 
(A) $d=4a$, (B) $d=10a$, (C) $d=30a$, (D)  $d=10000a$; 
where $a=0.246\, {\rm nm}$ is graphene's lattice constant,  
$\nu_a=1/8\pi\hbar va$ and $E_a=\hbar v/2a$.
(c)
LDOS vs. distance from the edge for different energies:
(A) $E=0.15E_a$, (B) $E=0.3E_a$, (C) $E=0.5E_a$. 
The delta function in Eq.~(\ref{LDOS}) is approximated 
by a Lorenzian $o/\pi(E^2+o^2)$ with $o=0.03E_a$.}
\label{LDOSfig}
\end{figure}

{\bf Introduction.}- 
Experimental evidence~\cite{Kostja05,Zhang05} 
for massless Dirac-like quasiparticles in graphene - 
a carbon monolayer with the hexagonal structure - 
has stimulated vigorous interest in electronic properties 
of this system 
(e.g. Refs.~\onlinecite{Gusynin05,Peres06,Brey06,McCann06,Nomura06,
Aleiner06,Tworzydlo06,Beenakker06,Cheianov07}). 
The unit cell of graphene contains two atoms 
each belonging to a triangular sublattice, 
and the low-energy states are described by a two-dimensional Dirac equation 
where the role of spin is assumed 
by the sublattice degree of freedom (pseudospin)~\cite{DiVincenzo84,Ando05}. 
Similar to relativistic spin-half particles in two dimensions, 
the graphene bulk density of states has a linear pseudogap~\cite{Kostja05} 
around zero energy $E=0$.
Natural boundaries can however give rise 
to additional spectral branches 
such as the low-energy edge states~\cite{Fujita96,Wakabayashi99}.
They are localized near a zigzag-shaped edge, 
whose outermost sites all belong to the same sublattice [Fig.~\ref{LDOSfig}(a)], 
and originate from the effective pseudospin "polarization" 
due to vanishing of one of the pseudospinor components as required by 
particle conservation~\cite{Brey06}.
Recent scanning tunneling experiments~\cite{Koba05,Niimi06} 
report a singular enhancement of the LDOS near zigzag boundaries 
attributed to the edge states. 

The measurements~\cite{Koba05,Niimi06} also revealed 
another peculiarity of the energy dependence of the LDOS -  
a fine oscillatory structure superimposed on the pseudogap 
with the amplitude enhanced at larger energies~\cite{Niimi06}. 
The origin of this behavior is still unaccounted for,  
although subsequent publications have studied the graphene LDOS,  
e.g. numerical simulations of Ref.~\onlinecite{Koba06} 
found damped spatial oscillations of the LDOS. 
The present study indends to show that both findings are   
consistent with the picture of interfering Dirac electron waves near a zigzag edge. 
To demonstrate this point, the one-particle Green's function 
of the Dirac equation was calculated for clean graphene with a zigzag edge described 
by the boundary condition of Ref.~\onlinecite{Brey06}. 
Then, the following expression for the LDOS  
$\nu(E,d)$, as a function of energy $E$ and distance $d$ from the edge, 
was obtained
\!
\begin{equation}
\nu(E,d)=|E|\frac{1+J_0\left(\frac{2Ed}{\hbar v}\right)}{\pi (2\hbar v)^2}-
\frac{ J_1\left(\left|\frac{2Ed}{\hbar v}\right|\right) }{ 4\pi\hbar v d }
+\frac{\delta(E)}{4\pi
 d^2}.
\label{LDOS}
\end{equation} 
Here the delta-functional term results from the dispersionless zero-energy edge state 
whereas the oscillating components given by the Bessel functions $J_0(2Ed/\hbar v)$ 
and $J_1(2|E|d/\hbar v)$ are due to interfering waves formed of the states 
belonging to the Dirac spectrum 
($v$ and $\hbar$ are the electron velocity and Planck's constant).  
For $2|E|d/\hbar v\gg 1$ the amplitude of the oscillations is 
proportional to $\sqrt{|E|/d}$ [see also Figs.~\ref{LDOSfig}(b) and (c)], 
which qualitatively agrees with both the experiment~\cite{Niimi06} 
and numerical simulations~\cite{Koba06}. 

Another issue this study focuses on is the 
connection between the local electronic structure 
of zigzag-terminated graphene and Fano scattering~\cite{Fano}. 
Unlike earlier works [e.g. Ref.~\onlinecite{Waka00}] where 
the Fano effect was due to resonant flux states in finite-size ribbons, 
here the Fano resonance is studied in a nanowire side-coupled
to half-infinite graphene, by analogy with similar quantum-dot structures~\cite{Kobayashi02,Johnson04,Fuhrer06,Miriam07},
and originates from a single dispersionless edge state. 
Also, unlike Ref.~\onlinecite{Waka00}, the main focus here is on  
the transport of correlated electrons in Josephson nanowires.  
The Fano effect is predicted to cause quite unusual behaviors
of the critical current such as enhancement by temperature 
and, under certain conditions, $0-\pi$ transitions 
similar to those in ferromagnetic Josephson junctions~\cite{Buzdin82,Ryazanov01,Kontos01,Bauer04,Golubov04,Buzdin05}.  
In the context of the Josephson effect in graphene nanostructures 
(e.g. Refs.~\onlinecite{Titov06,Zareyan06}) 
these issues have not yet been addressed.

{\bf Green's function of a zigzag ribbon.}- 
Assuming no scattering between the two valleys, $K$ 
and $K^\prime$, of graphene's Brillouin zone~\cite{Ando05}, 
one only needs to calculate the Green's function in one of them, 
e.g. $K$, where the Dirac equation reads    
$
[\sigma_0E + i\hbar v(\sigma_x\partial_x+\sigma_y\partial_y)]G=
\sigma_0\delta(x-x^\prime)\delta(y-y^\prime).	
$
Here the (retarded) Green's function matrix 
$G_{jk}$ with $j,k=A,B$, 
Pauli $\sigma_{x,y}$ and unity $\sigma_0$ matrices all act in pseudospin space. 
It suffices to solve the pair of equations for $G_{AA}$ and $G_{BA}$. 
After expanding in plane waves ${\rm e}^{ikx}$, 
the equations for the Fourier components are 
$
G_{BA|k}=(\hbar v/E)(k+\partial_y)G_{AA|k}$ 
and 
$
[\partial^2_y-q^2]G_{AA|k}=(E/\hbar^2 v^2)\delta(y-y^\prime)$
with $\quad q^2=k^2-(E/\hbar v)^2$. 
The solution can be sought in the form 
$G_{AA|k}(y,y^\prime)=a(y^\prime){\rm e}^{-qy}+b(y^\prime){\rm e}^{qy}
-E{\rm e}^{ -q|y-y^\prime| }/2\hbar^2 v^2q$,
where the last term is the Green's function of an unbounded system,  
and the coefficients $a(y^\prime)$ and $b(y^\prime)$ are to be found from the  
boundary conditions~\cite{Brey06} 
$G_{BA|k}|_{y=0}=(k+\partial_y)G_{AA|k}|_{y=0}=0$ and $G_{AA|k}|_{y=w}=0$.    
This yields the following result 
\!
\begin{eqnarray}
	&&
	G_{AA|k}(y,y^\prime)=E/2\hbar^2 v^2q
	\times
	\nonumber\\
	&&
	\times\left\{
	\frac{k[ \cosh q(w-|y-y^\prime|)-\cosh q(w-y-y^\prime) ]}{q\cosh qw
 -k\sinh qw}-\right.
	\nonumber\\
	&&
	\left.
	-\frac{q[ \sinh q(w-|y-y^\prime|)+\sinh q(w-y-y^\prime) ]}{q\cosh qw
 -k\sinh qw}
	\right\}.
	\label{G_k}
\end{eqnarray}
The poles of $G_{AA|k}$, given by the equation $q=k\tanh qw$ 
(cf. Ref.~\onlinecite{Brey06}), determine the excitation spectrum.
As known~\cite{Fujita96,Wakabayashi99}, 
it has an almost flat branch merging with the Fermi level $E=0$  
corresponding to a state exponentially decaying from the edge into the interior. 
This can be easily seen from Eq.~(\ref{G_k}) in the limit $w\to\infty$:   
\!
\begin{equation}
	G_{AA|k}(y,y^\prime)=-\frac{E{\rm e}^{-q|y-y^\prime|}}{2\hbar^2 v^2q}
	+\frac{(q+k)^2{\rm e}^{-q(y+y^\prime)}}{2qE}.
	\label{G_k_inf}
\end{equation}
The pole $E=0$ describes a dispersionless edge state existing for $k>0$. 
From Eq.~(\ref{G_k_inf}) an exact position representation for 
the Green's function 
$G_{AA}(xy,xy^\prime)=\int_{-\infty}^{\infty}dkG_{AA|k}(y,y^\prime)/(2\pi)$ 
can be obtained as
\!
\begin{eqnarray}
	&&
	G_{AA}(xy,xy^\prime)=
	\frac{EY_0(k_E|y-y^\prime|)-i|E|J_0(k_E|y-y^\prime|)}{(2\hbar v)^2}
	\nonumber\\
	&&
	+\frac{EY_0(k_E(y+y^\prime))-i|E|J_0(k_E(y+y^\prime))}{(2\hbar v)^2}
	\nonumber\\
	&&
	-\frac{2EY_1(k_E(y+y^\prime))-2i|E|J_1(k_E(y+y^\prime))}{(2\hbar
 v)^2k_E(y+y^\prime)},
	\label{G}
\end{eqnarray}
where $J_n(z)$ and $Y_n(z)$ ($n=0,1$) are, respectively, the Bessel and
Neumann functions, and $k_E=\sqrt{ E^2 }/\hbar v$. 
The LDOS [Eq.~(\ref{LDOS})] is obtained via 
$\nu(E,d)=-(1/\pi){\rm Im}G_{AA}(xy, xy^\prime)|_{y=y^\prime=d}$,
taking into account the pole of the function $Y_1$. 
We note that the interference of the waves 
incident at and reflected from the edge with small momenta $|k|\leq|E|/\hbar v$ 
produces spatial oscillations of the LDOS with the period much bigger than 
the lattice constant $a$ [Fig.~\ref{LDOSfig}(c)], i.e. 
well within the scanning tunneling microscop resolution.  
It is also instructive to examine Eq.~(\ref{G})
near the edge where it assumes a universal form, 
\!
\begin{equation}
G_{AA}(xy,xy^\prime)\approx 1/\pi E (y+y^\prime)^2, \qquad y,y^\prime\to 0,
\label{G1}
\end{equation}
independent of material parameters. 
To regulate the divergence at $y=y^\prime=0$, 
due to the effective continuum description,  
it is convenient to introduce the cutoff 
$G_{AA}(x0,x0)\approx 1/4\pi E d^2_c$ with $d_c\sim a$. 

\begin{figure}[t]
\epsfxsize=1\hsize
\epsffile{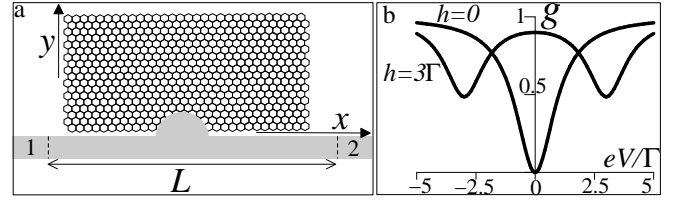}
\caption{
(a) Schematic view of a zigzag-edge graphene ribbon with 
a side tunnel contact to a nanowire
connecting electron reservoirs 1 and 2. 
The contact is assumed point-like, i.e. 
its size is much bigger than the interatomic distances, 
but smaller than the electronic mean free paths in both wire and graphene. 
(b) Zero-temperature conductance $g$ (in units of $e^2/\pi\hbar$) vs. 
bias voltage $V$ for spin-degenerate ($h=0$) and spin-split
($h=3\Gamma$) edge state in graphene. 
}
\label{Geo}
\end{figure}
   
{\bf Fano scattering off a zigzag edge.}-
The behavior of the Green's function near the edge 
can have a direct impact on charge transport.  
Let us consider a quasi-one-dimensional wire 
(with conventional quasiparticle spectrum) coupled in parallel
to a zigzag graphene edge via 
a point-like tunnel barrier [Fig.~\ref{Geo}(a)]. 
The contact is modelled by a real-space tunneling Hamiltonian of the form 
$H_T=\psi^\dagger_w(0)({\cal T}\psi_A({\bf r}_0)
  +{\cal T}^\prime\psi^\prime_A({\bf r}_0))+{\rm h.c.},\,
  {\bf r}_0=(0,0),$
where the electron operator in the wire $\psi^\dagger_w(0)$  
at the contact point $x=0$ is coupled to those in graphene on sublattice A in
both valleys $K$, $\psi_A({\bf r}_0)$ and  $K^\prime$, $\psi^\prime_A({\bf r}_0)$ 
with the matrix elements ${\cal T}$ and ${\cal T}^\prime$.   
To describe electron scattering caused by the contact, 
I use the equations-of-motion method and calculate 
the retarded Green's function in the wire 
\!
\begin{equation}
G_w(x,x^\prime)=G_w^{(0)}( x,x^\prime )+
	        \frac{G_w^{(0)}(x,0) \Sigma\, G_w^{(0)}(0,x^\prime)}
	             {1-G_w^{(0)}(0,0)\Sigma}.
\label{Gw}
\end{equation}
Here 
$G_w^{(0)}(x,x^\prime)={\rm e}^{ik_w|x-x^\prime|}/i\hbar v_F$ 
is the Green's function in the absence of tunneling 
($k_w\approx k_F + E/\hbar v_F$ with $v_F$ and $k_F$ 
being the Fermi velocity and wave number in the channel),   
$\Sigma =|{\cal T}|^2G_{AA}({\bf r}_0,{\bf r}_0)
         +|{\cal T}^\prime|^2G^\prime_{AA}({\bf r}_0,{\bf r}_0)$
is the tunneling self-energy, and 
$G^\prime_{AA}({\bf r}_0,{\bf r}_0)$ is the Green's function 
in valley $K^\prime$ coinciding with $G_{AA}({\bf r}_0,{\bf r}_0)$ [Eq.~(\ref{G1})].
The transmission amplitude~\cite{Fisher81} between the reservoirs is
$t=i\hbar v_F G_w\left(\frac{L}{2},-\frac{L}{2}\right)={\rm e}^{ik_wL}E/(E+i\Gamma)$,
where $\Gamma =(|{\cal T}|^2+|{\cal T}^\prime|^2)/4\pi d_c^2\hbar v_F$ 
determines the tunneling rate $\Gamma/\hbar$ between the systems.
The Fano-like transmission antiresonance at $E=0$
manifests complete backscattering of an electron wave incoming 
from one of the reservoirs.  
It is due to destructive interference 
between the electron wave directly transmitted through the wire 
(without tunneling) and the wave transmitted 
via tunneling through the graphene edge state whose energy 
is pinned to the Fermi level in the wire.   
It is straightforward to generalize the analysis 
to a spin-split edge state with energies 
$\mp h$ for spin projections $\alpha=\pm 1/2$. 
In this case, we have 
\!
\begin{equation}
t_\alpha(E)={\rm e}^{ik_wL}(E+2\alpha h)/(E+2\alpha h +i\Gamma).
\label{t}
\end{equation}
Figure~\ref{Geo}(b) shows the voltage dependence of the zero-temperature
Landauer conductance 
$g(V)=e^2/(2\pi\hbar)\sum_{\alpha=\pm 1/2}|t_\alpha(eV)|^2$. 
For $h\not =0$ the conductance dip is split due to 
the spin-filtering effect discussed earlier~\cite{Torio04,Lee06} 
in the context of possible applications in spintronics~\cite{Fabian04}. 

\begin{figure}[t]
\epsfxsize=0.6\hsize
\epsffile{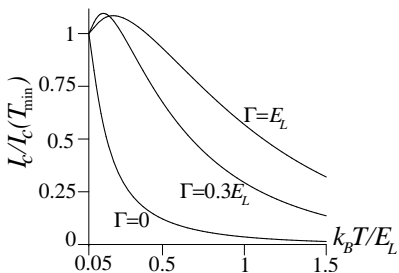}
\caption{
Critical current vs. temperature for a spin-degenerate edge state:
$h=0$, $\gamma=0.1E_L$, $\Delta=10E_L$. 
The current is normalized to the value $I_c(T_{min})$ 
where $T_{min}=0.05E_L/k_B$ is the lowest temperature for which 
the condition $\gamma\leq\pi k_BT\ll \Delta$ of weak proximity effect still holds.}
\label{IcT}
\end{figure}

{\bf Fano effect in a Josephson junction.}-
Let us finally discuss the case of superconducting reservoirs  
supporting an equilibrium Josephson current.  
The Josephson coupling is maintained
due to the Andreev process~\cite{Andreev} 
whereby an electron is retro-reflected as a Fermi-sea hole 
from one of the superconductors with the subsequent hole-to-electron 
conversion in the other one. 
Such an Andreev reflection circle facilitates a Cooper pair transfer 
between the superconductors.
Since both electron and hole also experience normal scattering inside the junction, 
the transmission antiresonance is expected 
to strongly influence the Josephson current. 
It is convenient to use the approach of Refs.~\onlinecite{Beenakker,Brouwer} 
relating the supercurrent to the scattering amplitudes 
via a sum over the Matsubara frequencies $\omega_n=(2n+1)\pi k_BT$ as follows
\begin{equation}
I_c=-4ek_BT/\hbar\sum_{n\geq 0,\alpha}
{\rm a}^2_{\alpha}(E)t_{\alpha}(E)t^*_{-\alpha}(-E)|_{E=i\omega_n},
\label{Ic}
\end{equation} 
where $t^*_{-\alpha}(-E)$ is the hole transmission amplitude 
corresponding to the time-reversed counterpart of the electron 
Hamiltonian~\cite{deGennes,Beenakker},
and ${\rm a}_{\alpha}(E)$ is the Andreev reflection amplitude at the
point contacts to superconductors 1 and 2. 
Equation~(\ref{Ic}) is applicable for arbitrary $t_{\alpha}(E)$ 
as long as ${\rm a}^2_{\alpha}$ is small enough so that 
one can neglect higher order Andreev processes. 
In this case the Josephson current-phase relation is sinusoidal 
with $I_c$ [Eq.~(\ref{Ic})] being the critical value of the current. 

\begin{figure}[t]
\epsfxsize=0.5\hsize
\epsffile{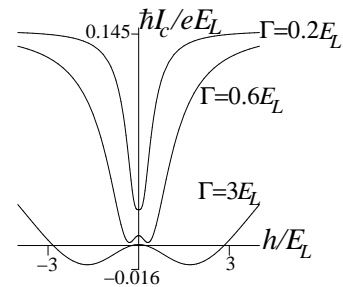}
\caption{
Critical current vs. spin-splitting energy:
$T=0.1E_L/k_B$, $\gamma=0.1E_L$, $\Delta=10E_L$.}
\label{IcH}
\end{figure}

In contacts to conventional Bardeen-Cooper-Schrieffer (BSC) superconductors, 
the Andreev process is well described by the scattering model of Ref.~\onlinecite{Blonder82}. 
However, in many practical cases superconducting contacts to low-dimensional 
systems can hardly be regarded as BCS-like ones. 
Proximity-effect contacts to 
semiconductor nanowires~\cite{vanDam,Tinkham} 
and carbon nanotubes~\cite{Jorgensen,Tsuneta} 
are important examples of such a situation. 
In this case a thin normal-metal layer is inserted between the superconductor and
the wire to ensure a good electrical contact.
In proximity-effect point contacts the Andreev scattering amplitude can be expressed 
in terms of the quasiclassical condensate ${\cal F}_{\alpha}(\omega_n)$ 
and quasiparticle ${\cal G}_{\alpha}(\omega_n)$ 
Green's functions of the normal layer as~\cite{GolKup,Volkov}
${\rm a}_{\alpha}(\omega_n)=i{\cal F}_{\alpha}/(1+{\cal G}_{\alpha})$.
I will adopt this approach and make use of McMillan's expressions~\cite{McMillan,McGolubov} 
for the Green's functions:
${\cal F}_{\alpha}=\Delta_n/\sqrt{\omega_n^2+\Delta_n^2}$, 
${\cal G}_{\alpha}=(\omega_n/\Delta_n){\cal F}_{\alpha}$,
and 
$\Delta_n=\gamma\Delta/(\gamma+\sqrt{\omega_n^2+\Delta^2})$,
where $\Delta$ is the superconductor's pairing energy and $\gamma$
is McMillan's parameter controlling the strength 
of the proximity effect in the normal layer and, hence, the Andreev
reflection amplitude 
${\rm a}_{\alpha}(\omega_n)=i\Delta_n/( \omega_n
 +\sqrt{\omega^2_n + \Delta^2_n} )$.  
For a weak proximity effect with $\gamma\leq\pi k_BT\ll\Delta$, 
the amplitude ${\rm a}^2_{\alpha}$ is small~\cite{Tkachov07} 
and equation~(\ref{Ic}) assumes the form
\!
\begin{equation}
I_c=\frac{8ek_BT}{\hbar}\sum\limits_{n\geq 0}
\frac{\Delta^2_n{\rm e}^{-\omega_n/E_L}}
{[\omega_n +\sqrt{\omega^2_n + \Delta^2_n} ]^2}
{\rm Re}\frac{( h+i\omega_n )^2}{[ h+i(\omega_n +\Gamma)  ]^2},
\nonumber
\end{equation} 
where the exponential factor results from the dynamical phase $2EL/\hbar v_F$
accumulated in the Andreev circle, introducing the Thouless energy  
$E_L=\hbar v_F/2L$.

\begin{figure}[t]
\epsfxsize=0.7\hsize
\epsffile{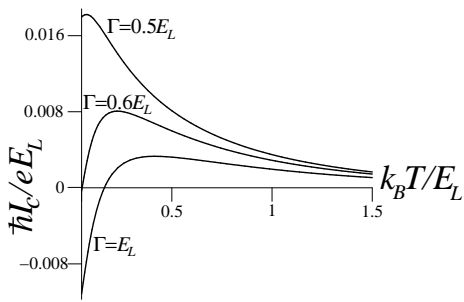}
\caption{
Critical current vs. temperature 
for a spin-polarized edge state:
$h=0.5E_L$, $\gamma=0.1E_L$, $\Delta=10E_L$.
}
\label{IcH_T}
\end{figure}

Figure~\ref{IcT} shows the temperature dependence of $I_c$
for the spin-degenerate case ($h=0$). 
In the absence of tunneling ($\Gamma=0$) it is just a monotonic
exponential decrease. However, for $\Gamma\not=0$ the interplay 
of the transmission antiresonance and the exponential suppression 
gives rise to a maximum at finite $T$.
Spin splitting of the graphene edge state lifts 
the Fano resonance condition $E=0$ for both electron 
and Andreev reflected hole. 
Therefore, for relatively weak tunneling coupling, 
when $\pi k_BT\leq\Gamma<E_L$,
the critical current increases with $h$ [Fig.~\ref{IcH}], 
a behavior quite unusual for Josephson junctions.   
Surprisingly, for stronger tunneling coupling ($\Gamma>E_L$) 
the function $I_c(h)$ becomes nonmonotonic with a rather broad region 
$h_1\leq |h|\leq h_2\approx \Gamma$ 
where $I_c$ is negative. 
The lower boundary $h_1\approx \pi k_BT$ is set by the temperature 
and is much smaller than all $E_L$, $\Gamma$ and $\Delta$. 
The supercurrent reversal is a consequence of the spin-dependent phases
acquired by both electron and hole due to scattering 
off the spin-polarized graphene edge, with   
negative values of $I_c$ implying a built-in $\pi$-phase difference 
in the ground state of a Josephson junction~\cite{Bulaevskii77} 
as opposed to $0$-phase difference for $I_c>0$. 
The $0-\pi$ transition can be driven by temperature as shown in Fig.~\ref{IcH_T}. 
Such a $\pi$ state is known to occur 
in ferromagnetic junctions 
where the condensate function oscillates 
in space~\cite{Buzdin82,Ryazanov01,Kontos01,Bauer04} 
(see, also recent reviews~\onlinecite{Golubov04,Buzdin05}). 
The author is not aware of any earlier work predicting $0-\pi$ transitions 
due to spin-dependent Fano scattering.
The main condition for this mechanism to work, i.e. $|h|>\pi k_BT$, 
can be met in the millikelvin region at modest external magnetic fields. 

The author thanks G. Cuniberti, M. Hentschel and C. Strunk for discussions. 
This work was partially funded by the European Union grant CARDEQ 
under Contract No. FP6-IST-021285-2. 




\begin{thebibliography}{00}
\bibitem{Kostja05}
K. S. Novoselov, A. K. Geim, S. V. Morozov, D. Jiang, M. I. Katsnelson,
I. V. Grigorieva, S. V. Dubonos, and A. A. Firsov, Nature (London)
{\bf 438}, 197 (2005).

\bibitem{Zhang05}
Y. Zhang, Y.-W. Tan, H. L. Stormer, and P. Kim, {\em ibid.} {\bf
 438}, 201 (2005).

\bibitem{Gusynin05}
V. P. Gusynin and S. G. Sharapov, Phys. Rev. Lett. {\bf 95}, 146801 (2005).
 
\bibitem{Peres06}
N. M. R. Peres, F. Guinea, and A. H. Castro Neto, Phys. Rev. B {\bf 73}, 125411 (2006).

\bibitem{Brey06}
L. Brey and H. A. Fertig, {\em ibid.} {\bf 73}, 235411 (2006). 


\bibitem{McCann06}
E. McCann and V. I. Falko, Phys. Rev. Lett. {\bf 96}, 086805 (2006).

\bibitem{Nomura06}
K. Nomura and A. H. MacDonald, {\em ibid.} {\bf 96}, 256602 (2006).

\bibitem{Aleiner06}
I. L. Aleiner and K. B. Efetov, {\em ibid.} {\bf 97}, 236801
 (2006).

\bibitem{Tworzydlo06}
J. Tworzydlo, B. Trauzettel, M. Titov, A. Rycerz, 
and C. W. J. Beenakker, 
{\em ibid.} {\bf 96}, 246802 (2006).

\bibitem{Beenakker06}
C. W. J. Beenakker, {\em ibid.} 97, 067007 (2006).

\bibitem{Cheianov07}
V.V. Cheianov, V. Fal'ko, B. L. Altshuler, Science {\bf 315}, 1252 (2007).


\bibitem{DiVincenzo84}
D. P. DiVincenzo and E. J. Mele, Phys. Rev. B {\bf 29}, 1685 (1984).

\bibitem{Ando05}
T. Ando, J. Phys. Soc. Jpn. {\bf 74}, 777 (2005).

\bibitem{Fujita96}
M. Fujita, K. Wakabayashi, K. Nakada, and K. Kusakabe, 
{\em ibid.} {\bf 65}, 1920 (1996);
K. Nakada, M. Fujita, G. Dresselhaus, and M. S. Dresselhaus, 
Phys. Rev. B {\bf 54}, 17954 (1996).

\bibitem{Wakabayashi99}
K. Wakabayashi, M. Fujita, H. Ajiki, and M. Sigrist, {\em ibid.} {\bf 59}, 8271 (1999).

\bibitem{Koba05}
Y. Kobayashi, K. I. Fukui, T. Enoki, K. Kusakabe, and Y. Kaburagi,
{\em ibid.} {\bf 71}, 193406 (2005).

\bibitem{Niimi06}
Y. Niimi, T. Matsui, H. Kambara, K. Tagami, M. Tsukada, and H.
Fukuyama, {\em ibid.} {\bf 73}, 085421 (2006).

\bibitem{Koba06}
Y. Kobayashi, K. I. Fukui, T. Enoki, and K. Kusakabe, {\em ibid.} {\bf 73}, 125415 (2006).

\bibitem{Fano}
U. Fano, Phys. Rev. {\bf 124}, 1866 (1961).

\bibitem{Waka00}
K. Wakabayashi and M. Sigrist, Phys. Rev. Lett. 84, 3390 (2000).

\bibitem{Kobayashi02}
K. Kobayashi, H. Aikawa, S. Katsumoto, and Y. Iye, {\em ibid.} {\bf 88}, 256806 (2002).

\bibitem{Johnson04}
A. C. Johnson, C. M. Marcus, M. P. Hanson, and A. C. Gossard, {\em ibid.} {\bf 93}, 106803 (2004).

\bibitem{Fuhrer06}
A. Fuhrer, P. Brusheim, T. Ihn, M. Sigrist, K. Ensslin, W. Wegscheider, and M. Bichler, 
Phys. Rev. B {\bf 73}, 205326 (2006).

\bibitem{Miriam07}
M. del Valle, R. Gutierrez, C. Tejedor, and G. Cuniberti,	
Nature Nanotechnology {\bf 2}, 176 (2007).


\bibitem{Buzdin82}
A. I. Buzdin, L. N. Bulaevskii, and S. V. Panyukov, 
Pis'ma Zh. Eksp. Teor. Fiz. {\bf 35}, 147 (1982) [JETP Lett. {\bf 35}, 178 (1982)].

\bibitem{Ryazanov01}
V. V. Ryazanov, V. A. Oboznov, A. Yu. Rusanov, A. V. Veretennikov, A. A. Golubov, and J. Aarts, 
Phys. Rev. Lett. {\bf 86}, 2427 (2001). 

\bibitem{Kontos01}
T. Kontos, M. Aprili, J. Lesueur, and X. Grison, {\em ibid.} {\bf 86}, 304 (2001).

\bibitem{Bauer04}
A. Bauer, J. Bentner, M. Aprili, M. L. Della Rocca, M. Reinwald, W. Wegscheider, and C. Strunk, 
{\em ibid.} {\bf 92}, 217001 (2004). 

\bibitem{Golubov04}
A. A. Golubov, M. Yu. Kupriyanov, and E. Il'ichev, Rev. Mod. Phys. {\bf 76}, 411 (2004).

\bibitem{Buzdin05}
A. I. Buzdin, {\em ibid.} {\bf 77}, 935 (2005).

\bibitem{Titov06}
M. Titov and C. W. J. Beenakker, 
Phys. Rev. B {\bf 74}, 041401(R) (2006).

\bibitem{Zareyan06}
Ali G. Moghaddam and Malek Zareyan, {\em ibid.} {\bf 74}, 241403(R) (2006).


\bibitem{Fisher81}
D. S. Fisher and P. A. Lee, {\em ibid.} {\bf 23}, 6851 (1981).

\bibitem{Torio04}
M. E. Torio, K. Hallberg, S. Flach, A. E. Miroshnichenko, and M. Titov, Eur. Phys. J. B {\bf 37}, 399 (2004).

\bibitem{Lee06}
M. Lee and C. Bruder, Phys. Rev. B {\bf 73}, 085315 (2006).

\bibitem{Fabian04}
I. \u{Z}uti\'{c}, J. Fabian, and S. Das Sarma, Rev. Mod. Phys. {\bf 76}, 323 (2004).



\bibitem{Andreev}
A.F. Andreev, Zh. Eksp. Teor. Fiz. {\bf 46}, 1823 (1964) 
[Sov. Phys. JETP {\bf 19}, 1228 (1964)].

\bibitem{Beenakker}
C.W.J. Beenakker, Phys. Rev. Lett. {\bf 67}, 3836 (1991);
in {\em Transport Phenomena in Mesoscopic Systems}, 
edited by H. Fukuyama and T. Ando, p. 235 (Springer, Berlin 1992).

\bibitem{Brouwer}
P.W. Brouwer and C.W.J. Beenakker, 
Chaos, Solitons and Fractals {\bf 8}, 1249 (1997).

\bibitem{deGennes}
P.G. de Gennes, Rev. Mod. Phys {\bf 36}, 225 (1964).

\bibitem{Blonder82}
G.E. Blonder, M. Tinkham, and T.M. Klapwijk, 
Phys. Rev. B {\bf 25}, 4515 (1982).


\bibitem{vanDam}
J. A. van Dam, Y. V. Nazarov, E. P. A. M. Bakkers, S. De Franceschi, L.
P. Kouwenhoven, Nature {\bf 442}, 667 (2006). 

\bibitem{Tinkham}
Jie Xiang, A. Vidan, M. Tinkham, R. M. Westervelt  
and Charles M. Lieber, Nature Nanotechnology {\bf 1}, 208 (2006).

\bibitem{Jorgensen}
H.I. Jorgensen, K. Grove-Rasmussen, T. Novotny, K. Flensberg,
and P.E. Lindelof, Phys. Rev. Lett. {\bf 96}, 207003 (2006).

\bibitem{Tsuneta}
T. Tsuneta, L. Lechner, and P. J. Hakonen, {\em ibid.} {\bf 98},
 087002 (2007).


\bibitem{GolKup}
A.A. Golubov and M.Yu. Kupriyanov, Physica C {\bf 259}, 27 (1996).

\bibitem{Volkov}
A. F. Volkov and A. V. Zaitsev, Phys. Rev. B {\bf 53}, 9267 (1996).

\bibitem{McMillan}
W. L. McMillan, Phys. Rev. {\bf 175}, 537 (1968).

\bibitem{McGolubov}
A. A. Golubov, E. P. Houwman, J. G. Gijsbertsen, V. M. Krasnov, J.
 Flokstra, H. Rogalla, and M. Yu. Kupriyanov, 
Phys. Rev. B {\bf 51}, 1073 (1995).

\bibitem{Tkachov07}
G. Tkachov and K. Richter, {\em ibid.} {\bf 75}, 134517 (2007),
discusses magnetic pair breaking as 
a means of controlling the Andreev amplitude in quantum dot Josephson junctions.

\bibitem{Bulaevskii77}
L. N. Bulaevskii, V. V. Kuzii, and A. A. Sobyanin, 
Pis'ma Zh. Eksp. Teor. Fiz. {\bf 25}, 314 (1977) 
[JETP Lett. {\bf 25}, 290 (1977)].

\end{thebibliography}
\end{document}